\documentclass[doublespacing]{elsart}
\usepackage{graphicx,amssymb}
\usepackage{epsf}
\usepackage{mathtext}
\newcommand{\erf}{\mathop{\rm erf}\nolimits}

\begin{document}
\begin{frontmatter}

\title{Coherent electron pair. Electromagnetic field}

\author[label1]{Constantin V. Usenko}, \author[label1]{Nataliia O. Cherkashyna\corauthref {cor}}
\corauth[cor]{Corresponding author}

\ead{ncherkashyna@ukr.net}

\address [label1] {National Taras Shevchenko University of Kyiv, Department of Theoretical Physics, Glushkova ave. 2, building 1, Kyiv 03127, Ukraine} 

\begin{abstract}
It is shown that relative coordinate and momentum of coherent electron pair
have the meaning of observables with the help of quadrupole and magnetic moments. Distributions of quadrupole terms of scalar potential are shown. These distributions have nonclassical properties.
\end{abstract}


\end{frontmatter}

\section{Introduction}

For many-particle systems correlation functions play the fundamental role for description of physical properties of entangled systems. 

The problem of such observable values as relative coordinate and momentum exists. The internal properties of electron pair become apparent in the electromagnetic interaction between the electrons, and between the electrons and their environment as well.

Recently the properties of separate quantum particles and few-particle quantum systems are investigated in view of quantum information theory. These properties are interesting because quantum gate can be realized by means of entangled electron pairs. Besides that, electron pairs with opposite spin propagate across periodic structures via direct and sequential two-electron tunneling. One of the most important characteristics of electrons scattering is the tunneling time, Ref. \cite{1,4,5,6,7,14,15,16}.
Entanglement shows the presence of non-classical correlations between quantum systems. For many-particle systems correlation functions play the fundamental role for description of physical properties of these systems. Thus, the investigation of relationship between the entanglement and correlation functions is very important, Ref. \cite{2}. 

The properties of entanglement are especially interesting in the case of continuous spectrum. One of the problems is the definition of relative coordinate and relative momentum in the tasks concerning systems of identical particles. The best example of this problem is the definition of relative coordinate and relative momentum of the coherent electron pair. 

The changeover to the c. m. system makes it possible to define the operators of relative coordinate and relative momentum: $\Psi \left( {x_1 ,x_2 } \right) = \pm \Psi 
\left( {x_2 ,x_1 } \right)$, $\Psi \left( {p_1 ,p_2 } \right) = \pm \Psi 
\left( {p_2 ,p_1 } \right)$. 

The average values of coordinate and momentum operators are zeroth: $\hat {x} = x_1 - x_2 $, $\hat {X} 
= \frac{x_1 + x_2 }{2}$, $\hat {p} = \frac{p_1 - p_2 }{2}$, $\hat {P} = p_1 
+ p_2 $.The average values of coordinate and momentum operators are: 
$\left\langle \hat {x} \right\rangle \equiv 0$,$\left\langle \hat {p} 
\right\rangle \equiv 0$.

Thus, we comes to the problem of such observable values as relative coordinate and momentum. The internal properties of electron pair become apparent in the electromagnetic interaction between the electrons, and between the electrons and their environment as well.  The observability of relative coordinates and momentums is investigated through the electromagnetic field of electron pair. This field is described by the value of vector-potential, but it's calculation is rather complicated. The effect of coordinate and momentum uncertainties is described through the magnetic moment and quadrupole moment of electron pair. The relative distance and relative velocity of classical electron pair are completely defined by of magnetic and quadrupole moments. The possibility to use this method for coherent electrons is the goal of our investigation and it is discussed below. 

One of the most important questions that are the subject of investigation today, is the research of spin-orbital interaction of entangled electrons.
    Investigation of spin-orbital interaction are at the top of scientific interest; they are at the top of the interest in individual cases as well, for example in approach in which space part of wave function of coherent electron pair is described by the Gaussian states. 
     Properties of the coherent state of a pair of almost free electrons are actively explored. The properties of systems that consist of a small number of charge bearers are under investigation. Special conditions at which Coulomb repulsion has small effect on the system and the energy of interaction can be considered as perturbation are found, see Ref. \cite{3,8,9}. 
     Theory of entanglement and properies of entanglement as well are very important, Ref. \cite{10}.
     Wave packages as a way of describing electrons and their behaviour, especially their complicated cases are investigted in Ref. \cite{11}.
     Entaglement is spin systems is also at the top of scientific interest, see Ref. \cite{12,13}.

\section {Electromagnetic field of single coherent electron}
In consideration of the electromagnetic field of coherent electron we use expressions for scalar and vector potentials of the electromagnetic field on determination from  electrostatics: 

\begin{equation}
\label{eq1}
\varphi (\vec{r})= e_0 \int {\frac{\rho \left( \vec {r'} \right)}{\left| \vec 
{r} - \vec {r'} \right|}d^3r'} ,
\end{equation}

\begin{equation}
\label{eq2}
\vec {A}(\vec {r}) = e_0 \int {\frac{j\left( \vec {r'} \right)}{\left|\vec {r} 
- \vec {r'} \right|}d^3r'} .
\end{equation}

Distributions of density of charge and current are determined by the wave function of electron.

\subsection {Wave function of single coherent electron}

Exploring the electromagnetic field of one electron (and electromagnetic field of coherent electron pair), we directed towards the investigation of effect of coordinate and momentum uncertainties of free particle. Here we see that a plane wave as an instrument for description of coherent electron state is not convenient. The same is true for functions that are localized in space, like Dirac function $\delta (x - x_0 (t))$. Instead of such wave functions, we consider a superposition of plane waves:

\begin{equation}
\label{eq3}
\Psi (x) = \frac{1}{\sqrt {2\pi \hbar } }\int {\Psi (p)\exp {\left( 
{i\frac{px}{\hbar }} \right) }dp }
\end{equation}

Probability amplitudes $\Psi (x)$ and $\Psi (p)$ give distributions of coordinate and momentum probabilities, respectively:

\begin{equation}
\label{eq4}
\rho (\vec {r},\vec {r}_0 ,\vec {p}_0 ) = e_0 \left| {\Psi (\vec {r},\vec 
{r}_0 ,\vec {p}_0 )} \right|^2,
\end{equation}

\begin{equation}
\label{eq5}
\begin{array}{l}
\vec {j}(\vec {r},\vec {r}_0 ,\vec {p}_0 ) = \\
 =  \frac{e_0 }{mc} \cdot 
\frac{1}{2}\left[{\Psi ^\ast (\vec {r},\vec {r}_0 ,\vec {p}_0 )\nabla \Psi 
(\vec {r},\vec {r}_0 ,\vec {p}_0 ) + \Psi (\vec {r},\vec {r}_0 ,\vec {p}_0 
)\nabla \Psi ^\ast (\vec {r},\vec {r}_0 ,\vec {p}_0 )}\right].
\end{array}
\end{equation}

Such distributions of probabilities are characterized by the coordinate and 
momentum uncertainties, $\sigma _x $ and $\sigma _p $. Using the Heisenberg relation of uncertainties, which has minimum for the coherent states in the certain moment of time, we have:

\begin{equation}
\label{eq6}
\sigma _x \sigma _p = \frac{\hbar }{2}.
\end{equation}

For a free particle the momentum uncertainty is constant $\sigma _p = 
const$, but the coordinate uncertainty depends on time. The equality (\ref{eq4}) is changed to inequality because of wave package spreading with time. We suppose that the coordinate and momentum uncertainties are the same for all directions.

Taking into account properties of the parameters of wave function, which are discussed above, we have the wave function of coherent electron: 
\begin{equation}
\label{eq7}
\begin{array}{l}
\Psi (\vec {r},\vec {r}_0 ,\vec {p}_0 ) = \\
=\frac{1}{\left( {\sqrt {\sigma 
\sqrt {2\pi (1 + \omega ^2(t - t_0 )^2)} } } \right)^3}\exp{ \left( { - 
\frac{\left( {\vec {r} - \vec {r}_0 - \frac{\vec {p}_0 }{m}(t - t_0 )} 
\right)^2}{4\sigma ^2(1 + \omega ^2(t - t_0 )^2)} + i\frac{\vec {p}_0 \vec 
{r}}{\hbar }} \right)},
\end{array}
\end{equation}

where $\sigma $ is initial coordinate uncertainty, $r_0 $ is an average value of initial coordinate of electron,  $p_0 $- an average value of momentum of electron,  $t_0 $ is culmination moment (the instant when correlation between coordinate and momentum is absent). 

We put the coordinate origin to the point, which is described by radius vector $\vec {r}_0 $ and  rewrite the wave function of coherent electron as: 

\begin{equation}
\label{eq8}
\begin{array}{l}
\Psi (\vec {r},\vec {p}_0 ) = \\
= \frac{1}{\left( {\sqrt {\sigma \sqrt {2\pi (1 
+ \omega ^2(t - t_0 )^2)} } } \right)^3}\exp{ \left( { - \frac{\left( {\vec 
{r} - \frac{\vec {p}_0 }{m}(t - t_0 )} \right)^2}{4\sigma ^2(1 + \omega ^2(t 
- t_0 )^2)} + i\frac{\vec {p}_0 \vec {r}}{\hbar }} \right)},
\end{array}
\end{equation}

In the culmination moment ($t = t_0 $), we have simplified case of wave 
function of coherent electron:

\begin{equation}
\label{eq9}
\Psi (\vec {r},\vec {p}_0 ) = \frac{1}{\left( {\sqrt {\sigma \sqrt {2\pi } } 
} \right)^3}\exp{\left( { - \frac{\vec {r}^2}{4\sigma ^2} + i\frac{\vec 
{p}_0 \vec {r}}{\hbar }} \right)}. 
\end{equation}

The average value of momentum $\vec {p}_0 $ for the free particle 
characterize it's velocity $\vec {V} = \frac{\vec {p}_0 }{m}$. The velocity 
determines the vector-potential of electromagnetic field. 

\subsection{Potentials of electromagnetic field of single coherent electron}

In the consideration of electromagnetic field of coherent electron, we use expressions for scalar and vector potentials of the electromagnetic field
(\ref{eq1}),(\ref{eq2}). 

Densities of charge and current are: 

\begin{equation}
\label{eq10}
\rho (\vec {r}) = \frac{1}{\left( {\sqrt {\sigma \sqrt {2\pi } } } 
\right)^3}\exp{ \left( { - \frac{\vec {r}^2}{4\sigma ^2}} \right)}, 
\end{equation}

\begin{equation}
\label{eq11}
j(\vec {r}) = \frac{\vec {p}_0 }{mc}\rho (\vec {r}).
\end{equation}

Thus we obtain:

\begin{equation}
\label{eq12}
\varphi (\vec{r}) = e_0 \frac{1}{\sqrt 2 \sigma }Na\left( \vec{r} \right),
\end{equation}

\begin{equation}
\label{eq13}
\vec {A}(\vec{r}) = \frac{\vec {p}_0 }{mc}\varphi (\vec{r}),
\end{equation}

$Na\left( x \right)$ is defined in Appendix \ref{App}.

The same result as (\ref{eq10}) and (\ref{eq11}) can be obtained by application of Lorentz transformations to the field of static charge in the system which moves with the velocity  $\frac{\vec {p}_0 }{m}$. In special case, when $p_0 = 0$, 
density of current takes zero value. That is why the vector-potential of electromagnetic field is zeroth as well $\vec {A}(\vec{r}) = 0$. 

\begin{figure}
\begin{center}
\includegraphics*[width=5cm]{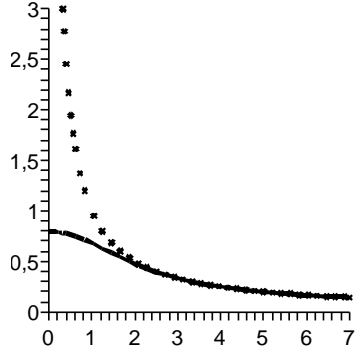}
\end{center}
\caption{Potential of electromagnetic field of coherent electron.}
\label{fig:1}
\end{figure}

In Fig. \ref {fig:1} dependence of scalar potential of one electron on distance between the center of charge and point of observation of potential is shown.  Sequence of points corresponds to the case of classic Coulomb potential, and line corresponds to potential of coherent electron. 

\break

Potential of electromagnetic field of coherent electron does not tend to 
infinity, as it is obvious from the Fig. \ref {fig:1}. It amounts to constant. This is 
explained in the following way. Error integral  $\erf\left(x\right)$ tends to 
1 if $x$ is more then 1: 

\begin{equation}
\label{eq14}
\erf\left( x \right) \cong 1 - \frac{1}{2\sqrt \pi }\frac{1}{x}\exp{ ( - x^2)}.  
\end{equation}

Till $x$ is big enough $\erf\left( x \right)$ is well calculated.

The second term of expansion is negligibly small because of $\exp{ ( - 
x^2)}$(as well as all following terms of expansion), that is why $\erf\left( 
x \right) = 1$. The scalar and vector potentials are: 

\begin{equation}
\label{eq15}
\varphi (\vec{r}) = \frac{e_0 }{\left| \vec {r} \right|},
\end{equation}

\begin{equation}
\label{eq16}
\vec {A}(\vec{r}) = \frac{\vec {p}_0 }{mc}\frac{e_0 }{\left| \vec {r} \right|}. 
\end{equation}

In the area which is smaller than coordinate uncertainty, the probability of registration of electron is small. That is why the field, produced by the electron, becomes smaller too. 
Thus, field of coherent electron only outside an area being few coordinate uncertainties coincides with the field of point charge.

\section {Electromagnetic field of coherent electron pair }

The value of charge density of electron pair is the sum of charge densities of both electrons:

\[
\rho (\vec {r}) = e_0 \int {\left| {\Psi \left( {\vec{r_1} , \vec{r_2} } \right)} 
\right| ^2\left( {\delta \left( {\vec {r} - \vec {r}_1 } \right) + \delta 
\left( {\vec {r} - \vec {r}_2 } \right)} \right)d^3r_1 d^3r_2} ,
\]

$\delta \left( {\vec {r} - \vec {r}_1 } \right)$describes the contribution of one electron, $\delta \left( {\vec {r} - \vec {r}_2 } 
\right)$ describes the contribution of another one.

The definition of current density is a similar sum of current densities of both electrons. 
 
\subsection {Wave function of coherent electron pair}

A wave function of coherent electron pair is symmetric or antisymmetric combination of wave functions of single electrons.  In the case when electron spins have antisymmetric mutual orientation wave function must be symmetric combination, and in the case when spins are parallel wave function must be antisymmetric combination (in accordance to Pauli principle).

\begin{equation}
\label{eq17}
\Psi (\vec {r}_1 ,\vec {r}_2 ) = \frac{\Psi _1 (\vec {r}_2 )\Psi _2 (\vec 
{r}_1 )\pm \Psi _1 (\vec {r}_1 )\Psi _2 (\vec {r}_2 )}{\sqrt {1\pm N^2} }, 
\end{equation}

where $N$ is overlapping integral, which is determined as: 

\begin{equation}
\label{eq18}
N = \int {\Psi _1^\ast (\vec{r})} \Psi _2 (\vec{r})d^3r = \left\langle {\Psi _1 (\vec{r})} 
\mathrel{\left| {\vphantom {{\Psi _1 (\vec{r})} {\Psi _2 (\vec{r})}}} \right. 
\kern-\nulldelimiterspace} {\Psi _2 (\vec{r})} \right\rangle .
\end{equation}

Substitution of (\ref{eq6}) to (\ref{eq17}) gives an explicit form of the wave function of coherent electron pair. 
Average values of coordinates are:

\[
\left\langle {\vec {r}_1 } \right\rangle + \left\langle {\vec {r}_2 } 
\right\rangle = \left\langle {\vec {r}_1 + \vec {r}_2 } \right\rangle .
\]
This is the average value of centre of mass coordinate in classic mechanics. 
We choose the reference frame which coincides with the c. m. system. Consequently, 
 $\vec {r}_{01} = \vec {r}_0 $, $\vec {r}_{02} = - \vec {r}_0 
$, $\vec {p}_{01} = - \vec {p}_0 $, $\vec {p}_{01} = \vec {p}_0 $. The wave 
function is represented as:

\begin{equation}
\label{eq19}
\begin{array}{l}
 \Psi (\vec {r}_1 ,\vec {r}_2 ) = \frac{1}{\sqrt {1\pm N^2} }\frac{1}{\left( 
{\sigma \sqrt {2\pi (1 + \omega ^2(t - t_0 )^2)} } \right)^3} \cdot \\ 
 \cdot \left[ {\exp \left( { - \frac{\left( {\vec {r}_1 - \vec {r}_0 - 
\frac{\vec {p}_0 }{m}(t - t_0 )} \right)^2}{4\sigma ^2(1 + \omega ^2(t - t_0 
)^2)} - \frac{\left( {\vec {r}_2 + \vec {r}_0 - \frac{\vec {p}_0 }{m}(t - 
t_0 )} \right)^2}{4\sigma ^2(1 + \omega ^2(t - t_0 )^2)} + i\frac{\vec {p}_0 
(\vec {r}_1 - \vec {r}_2 )}{\hbar }} \right)} \right.\pm \\ 
 \pm \left. {\exp \left( { - \frac{\left( {\vec {r}_2 - \vec {r}_0 - 
\frac{\vec {p}_0 }{m}(t - t_0 )} \right)^2}{4\sigma ^2(1 + \omega ^2(t - t_0 
)^2)} - \frac{\left( {\vec {r}_1 + \vec {r}_0 - \frac{\vec {p}_0 }{m}(t - 
t_0 )} \right)^2}{4\sigma ^2(1 + \omega ^2(t - t_0 )^2)} - i\frac{\vec {p}_0 
(\vec {r}_1 - \vec {r}_2 )}{\hbar }} \right)} \right] \\ 
 \end{array},
\end{equation}

where $N = \exp \left( { - \frac{2p_0^2 \sigma ^2}{\hbar ^2} - \frac{r_0^2 
}{2\sigma ^2}} \right)$. 

We want to prove that parameters $\vec {r}_0 $ and $\vec {p}_0 $ have the meaning of relative coordinate and relative momentum of coherent electron pair, respectively. The discussed problem is that the parameters $\vec 
{r}_0 $ and $\vec {p}_0 $ of two-particle function can not be considered as average values of observable operators of relative coordinate and relative momentum. The reason is that average values turn to zero as a result of identity of electrons. 

Now we consider coherent electron pair in the moment of culmination, $t = t_0 $. Then 
two-particle wave function is: 
\begin{equation}
\label{eq20}
\begin{array}{l}
\Psi (\vec {r}_1 ,\vec {r}_2) = \frac{1}{\sqrt {1\pm N^2} }\frac{1}{\left( 
{\sigma \sqrt {2\pi } } \right)^3} \cdot \\ 
 \cdot \left[ {\exp \left( { - \frac{\left( {\vec {r}_1 - \vec {r}_0 } 
\right)^2}{4\sigma ^2} - \frac{\left( {\vec {r}_2 + \vec {r}_0 } 
\right)^2}{4\sigma ^2} + i\frac{\vec {p}_0 (\vec {r}_1 - \vec {r}_2 )}{\hbar 
}} \right)} \right.\pm \\
\pm \left. {\exp \left( { - \frac{\left( {\vec {r}_2 - 
\vec {r}_0 } \right)^2}{4\sigma ^2} - \frac{\left( {\vec {r}_1 + \vec {r}_0 
} \right)^2}{4\sigma ^2} + i\frac{\vec {p}_0 (\vec {r}_1 - \vec {r}_2 
)}{\hbar }} \right)} \right]. \\ 
\end{array}
\end{equation}

\subsection {Potentials of electromagnetic field of coherent electron pair}

We calculate scalar and vector potentials of the electromagnetic field by (\ref{eq6}) and (\ref{eq7}). The potential is determined by (\ref{eq8}).
We have explicit form of charge density:  

\begin{equation}
\label{eq21}
\begin{array}{l}
 \rho (\vec {r}) = \frac{e_0}{1 + N^2}\exp \left( { - \frac{(\vec {r} - 
\vec {r}_0 )^2}{2\sigma ^2}} \right) + \exp \left( { - \frac{(\vec {r} + 
\vec {r}_0 )^2}{2\sigma ^2}} \right) + \\ 
 + \exp \left( { - \frac{r^2 + 2r_0^2 }{2\sigma ^2}} \right)\exp \left( { - 
\frac{2p_0^2 \sigma ^2}{\hbar ^2}} \right)\left[ {\exp \left( {\frac{2i\vec 
{p}_0 \vec {r}}{\hbar }} \right) + \exp \left( { - \frac{2i\vec {p}_0 \vec 
{r}}{\hbar }} \right)} \right]. \\ 
 \end{array}
\end{equation}

We consider the case of central impact of coherent electrons. In this case 
$\vec {p}_0 $ is parallel to $\vec {r}$ According to the formulae (\ref{eq8}) and (\ref{eq9}) the values of scalar potential of coherent electron pair are the next:

\begin{equation}
\label{eq22}
\begin{array}{l}
 \varphi_s (\vec{r}) = \frac{e_0 }{1 + \exp \left( { - \frac{4p_0^2 \sigma 
^2}{\hbar ^2} - \frac{r_0^2 }{\sigma ^2}} \right)}\left[ {\frac{1}{\sqrt 2 
\sigma }Na(\vec{r} - \vec{r}_0 ) + } \right.\frac{1}{\sqrt 2 \sigma }Na(\vec{r} + \vec{r_0} ) + \\ 
 + \exp \left( { - \frac{4p_0^2 \sigma ^2}{\hbar ^2} - \frac{r_0^2 }{\sigma 
^2}} \right)\left\{ {\frac{1}{\sqrt 2 \sigma }Na\left( {\vec{r} + \frac{2i\vec{p}_0 \vec{r}
\sigma ^2}{\hbar }} \right)\exp \left( {\frac{2i\vec{p}_0\vec{r} }{\hbar }} \right) + } 
\right. \\ 
 \left. {\left. { + \frac{1}{\sqrt 2 \sigma }Na\left( {\vec{r} - \frac{2i\vec{p}_0\vec{r} 
\sigma ^2}{\hbar }} \right)\exp \left( { - \frac{2i\vec{p_0}\vec{r} }{\hbar }} \right)} 
\right\} } \right], \\ 
 \end{array}
\end{equation}

\begin{equation}
\label{eq23}
\begin{array}{l}
 \varphi _a (r) = \frac{e_0 }{1 - \exp \left( { - \frac{4p_0^2 \sigma 
^2}{\hbar ^2} - \frac{r_0^2 }{\sigma ^2}} \right)}\left[ {\frac{1}{\sqrt 2 
\sigma }Na(\vec{r} - \vec{r}_0 ) + } \right.\frac{1}{\sqrt 2 \sigma }Na(\vec{r} + \vec{r}_0 ) - \\ 
 - \exp \left( { - \frac{4p_0^2 \sigma ^2}{\hbar ^2} - \frac{r_0^2 }{\sigma 
^2}} \right)\left\{ {\frac{1}{\sqrt 2 \sigma }Na\left( {\vec{r} + \frac{2i\vec{p}_0\vec{r} 
\sigma ^2}{\hbar }} \right)\exp \left( {\frac{2i\vec{p}_0\vec{r} }{\hbar }} \right) + } 
\right. \\ 
 \left. {\left. { + \frac{1}{\sqrt 2 \sigma }Na\left( {\vec{r} - \frac{2i\vec{p}_0\vec{r} 
\sigma ^2}{\hbar }} \right)\exp \left( { - \frac{2i\vec{p_0}\vec{r} }{\hbar }} \right)} 
\right\} } \right], \\ 
 \end{array}
\end{equation}

where $\varphi _s (r)$ is the scalar potential of the coherent electron pair with antiparallel mutual spin orientation;  $\varphi _a (r)$ is the scalar potential of the coherent electron pair with parallel mutual spin orientation. 

\begin{figure}
\begin{center}
\includegraphics*[width=5cm]{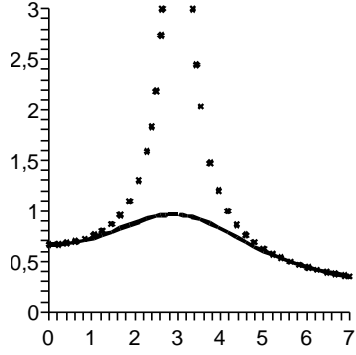}
\end{center}
\caption{Potential of electromagnetic field of coherent electron pair.}
\label{fig:2}
\end{figure}

In the Fig. \ref {fig:2} the line corresponds to the potential of the electrons with parallel spins and that of electrons with antiparallel spins as well (for the large values of relative coordinate they coincide). Sequence of points corresponds to the potential of electromagnetic field, produced by the pair of classical electrons. 

The calculations make evidence that vector potentials of fields produced by the coherent electron pair with antiparallel and parallel mutual spin orientation are:

\begin{equation}
\label{eq24}
\vec {A}_s (\vec{r}) = \frac{\vec {p}_0 }{mc}\varphi _s \left( \vec{r} \right)  
\end{equation}

\begin{equation}
\label{eq25}
\vec {A}_a (\vec{r}) = \frac{\vec {p}_0 }{mc}\varphi _a \left( \vec{r} \right)  
\end{equation}

\section{Moments of electromagnetic field produced by coherent electron pair}

We use the magnetic and quadrupole moments of electromagnetic field because the effect of coordinate and momentum uncertainties on electromagnetic field is investigated through the magnetic moment and quadrupole moment of electron pair. The relative distance and relative velocity of classical electron pair is fully defined with the help of magnetic and quadrupole moments. We don't use dipole moment of electromagnetic field of electron pair because it is equal to zero for the systems which consist of particles with the same specific charges. 
Relative coordinate determines the electric quadrupole moment of classical electron pair, relative momentum determines the magnetic moment of them. This dependence remains also true for the case of quantum particles, if the relative coordinate and momentum are large enough compared to their uncertainties.
A scalar potential of electromagnetic field is represented as sequence of multipole terms:

\begin{equation}
\label{eq26}
\varphi (\vec{r}) = \varphi ^{(0)} + \varphi ^{(1)} + \varphi ^{(2)},
\end{equation}

where 
\begin{itemize}
	\item[\ ]
$\varphi ^{(0)} =Q/r$ is potential of field of charge;

\item[\ ] $\varphi ^{(1)} = \vec {p}\cdot\vec {r}/r^3,$ is potential of
field of dipole;

\item[\ ] $\varphi ^{(2)} = \sum {D_{\alpha ,\beta }x^\alpha x^\beta} /r^5$
is potential of field of quadrupole;
\end{itemize}

\begin{equation}
\label{eq27}
D_{\alpha \beta } = \sum\limits_i {q_i  \left( {3x_i^\alpha x_i^\beta - 
r_i^2 \delta _{\alpha \beta } } \right)} .
\end{equation}

In the case of electron pair $\varphi ^{(1)} = 0$ because of the identity of electron and scalar potential of electromagnetic field consists of coulomb and quadrupole terms:

\[
\varphi (\vec{r}) = \varphi ^{(0)} + \varphi ^{(2)}.
\]

In the case if the one observes the field at the large distance from the system of charges  $r > > r_0 $, or $r > > p_0 \frac{\sigma ^2}{\hbar }$, 
scalar potential is represented as:

\[
\varphi (\vec{r}) = \frac{2e_0 }{\vec{r}} + D_{\alpha \beta } \frac{n^\alpha n^\beta 
}{r^3}.
\]

Vector potential produced by magnetic moment is represented as: 

\begin{equation}
	\vec {A}=\nabla \frac{1}{R}_0 \vec {m},
\end{equation}

where $ {R}_0 $ is distance from the charge to the point of observation, 

\[
\vec {m} = \frac{e_0 }{2c}\left[ {\vec {r}_0 \times \vec {V}_0 + \vec {r}_0 
\times \vec {V}_0 } \right] = \frac{e_0 }{mc}\vec {r}_0 \times \vec {p}_0
\]
is classical  value of magnetic moment of pair.

\subsection {Magnetic moment of magnetic field produced by the coherent 
electron pair }

If the system is quantum, for determination of correlation between the observables values we calculate quantum-mechanic average value of operator of magnetic moment:

\[
\left\langle \vec {m} \right\rangle = \frac{e_0 }{2c}\int {\Psi \ast (\vec 
{r}_1 ,\vec {r}_2 )\left[ {\vec {r}_1 \times \vec {V}_1 + \vec {r}_2 \times 
\vec {V}_2 } \right]\Psi (\vec {r}_1 ,\vec {r}_2 )d^3r_1 } d^3r_2 ,
\]

where $\Psi (\vec {r}_1 ,\vec {r}_2 )$ - wave function of coherent electron 
pair.

Using wave function of coherent electron pair we obtain average value of magnetic moment: 

\[
\left\langle \vec {m} \right\rangle = - \frac{e_0}{cm}\frac{1}{1 \mp \left| 
N \right|^2}\vec {r}_0 \times \vec {p}_{0x} ,
\]

where $N$ is overlapping integral, signs - or + in denominator correspond to the cases of parallel and antiparallel mutual spin orientation, respectively. 

Quantum properties of the system become apparent at  only. If a value of overlapping integral tends to the zero, the average value of magnetic moment tends to a value, which is specific for a pair of classic electrons. The effect of magnetic moment is essential on small distances from the system of charges only.

\subsection{Quadrupole moment of electromagnetic field produced by the 
coherent electron pair }

In the case of arbitrary mutual orientation of vectors  $\vec {p}_0 $ and $\vec 
{r}$ ($\vec {r} = (x,y,z)$, we choose direction of axes $\vec {p}_0 = 
(p_{0x} ,0,p_{0z} )$, $\vec {r}_0 = (0,0,r_0 )$), and calculate tensor components of quadrupole moment. 

\begin{equation}
\label{eq28}
D = \left( {{\begin{array}{*{20}c}
 {D_{xx} } \hfill & 0 \hfill & {D_{xz} } \hfill \\
 0 \hfill & {D_{yy} } \hfill & 0 \hfill \\
 {D_{zx} } \hfill & 0 \hfill & {D_{zz} } \hfill \\
\end{array} }} \right),
\end{equation}

where $D_{xz} = D_{zx} $.

The nonzero tensor components are defined as: 

\begin{equation}
\label{eq29}
D_{xx} = \int {\rho \left( {x,y,z} \right)\left( {2x^2 - y^2 - z^2} \right) 
dxdydz},
\end{equation}

\begin{equation}
\label{eq30}
D_{yy} = \int {\rho \left( {x,y,z} \right)\left( {2y^2 - x^2 - z^2} \right) 
dxdydz},
\end{equation}

\begin{equation}
\label{eq31}
D_{zz} = \int {\rho \left( {x,y,z} \right)\left( {2z^2 - x^2 - y^2} \right) 
dxdydz},
\end{equation}

\begin{equation}
\label{eq32}
D_{xz} = D_{zx} = \int {\rho \left( {x,y,z} \right)xz\ dxdydz},
\end{equation}

where $\rho (x,y,z)$ is a charge density.

An explicit form of tensor components of quadrupole moment (in the frame which is related to coherent electron pair' frame) is written down as:

\begin{equation}
\label{eq33}
D_{xx} = \frac{2N^2\left( {\frac{4\sigma ^4}{\hbar ^2}\left( {p_{0z}^2 - 
2p_{0x}^2 } \right)} \right) - 2r_0^2 }{1\pm N^2},
\end{equation}

\begin{equation}
\label{eq34}
D_{yy} = \frac{2N^2\left( {\frac{4\sigma ^4}{\hbar ^2}\left( {p_{0z}^2 + 
p_{0x}^2 } \right)} \right) - 2r_0^2 }{1\pm N^2},
\end{equation}

\begin{equation}
\label{eq35}
D_{zz} = \frac{2N^2\left( {\frac{4\sigma ^4}{\hbar ^2}\left( {p_{0x}^2 - 
2p_{0z}^2 } \right)} \right) + 4r_0^2 }{1\pm N^2},
\end{equation}

\begin{equation}
\label{eq36}
D_{xz} = D_{zx} = \frac{ - 12N^2\frac{\sigma ^2p_{0x} p_{0z} }{\hbar 
^2}}{1\pm N^2}.
\end{equation}

Let us consider the case $N \to 0$. It becomes possible if $\frac{4(p_{0x}^2 
+ p_{0z}^2 )\sigma ^2}{\hbar ^2} > > 1$ and $ - \frac{r_0^2 }{\sigma ^2} > > 
1$. That is $r_0 > > \sigma $ and $(p_{0x}^2 + p_{0z}^2 ) > > \sqrt 
{\frac{\hbar }{4\sigma }} $. 

In this case (\ref{eq33})-(\ref{eq36}) we obtain:

\[
D_{zz} = - 2D_{xx} = - 2D_{yy} = 4e_0 r_0^2.
\]

We get the following expressions for observable $r_0 $ by the quadrupole 
moment tensor components: 

\begin{equation}
\label{eq37}
r_0 = \frac{1}{2}\sqrt {\frac{D_{zz} }{e_0 }} .
\end{equation}

In order to define observables $p_{0x} $ and $p_{0z} $, let us consider the 
opposite case $N \to 1$. It is possible if $\frac{4(p_{0x}^2 + p_{0z}^2 
)\sigma ^2}{\hbar ^2} < < 1$ and $ - \frac{r_0^2 }{\sigma ^2} < < 1$, that 
is $p_0 < < \frac{\hbar }{2\sigma }$ and $r_0 < < \sigma $.

Under these conditions for observables $p_{0x} $ and $p_{0z} $ from (\ref{eq33})-(\ref{eq36}) we obtain following expressions:

\begin{equation}
\label{eq38}
p_{0x} = \frac{\hbar }{2\sigma ^2}\sqrt { - \frac{D_{zz} + 2D_{xx} }{3e_0 }} 
,
\end{equation}

\begin{equation}
\label{eq39}
p_{0z} = \frac{1}{\hbar }\sqrt {\frac{D_{xz} }{3e_0 \sqrt { - (D_{zz} + 
2D_{xx} )} }} .
\end{equation}

\begin{figure}

\begin{center}
    \includegraphics*[width=5cm] {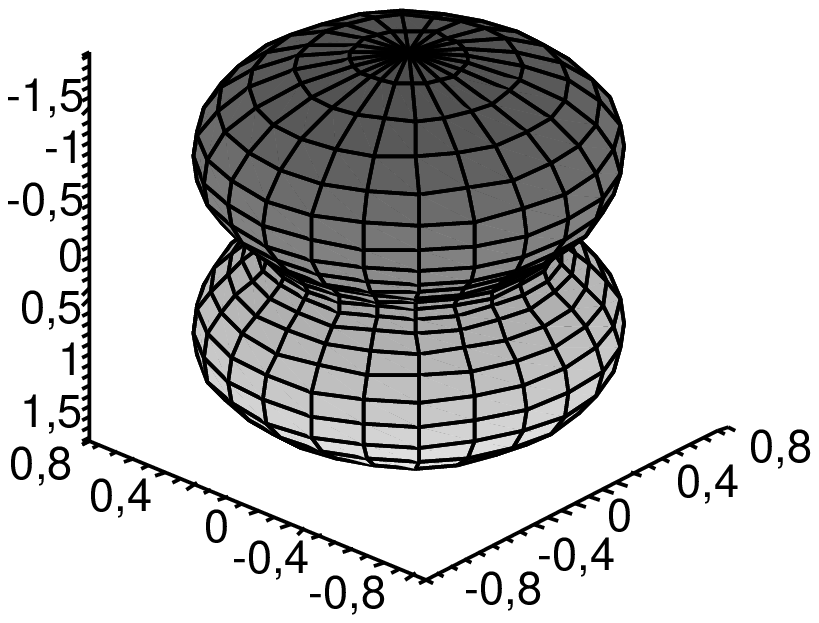}
    \end{center}
    \caption {A surface of quadrupole term of scalar potential,$\vec {r}_0 \vert \vert 
\vec {p}_0 $, $r_0 = 0.42$, $p_0 = 0.23$. }
\label {fig:4}
\end{figure}

\begin{figure}
\begin{center}
    \includegraphics*[width=5cm]{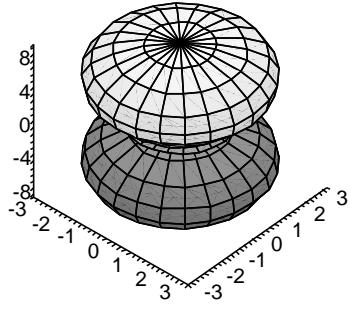} 
    \end{center}
    \caption {A surface of quadrupole term of scalar potential, $r_0 = 0.42$, $p_0 = 0$. }
\label {fig:5}

\end{figure}

\begin{figure}

\begin{center}
    \includegraphics*[width=5cm]{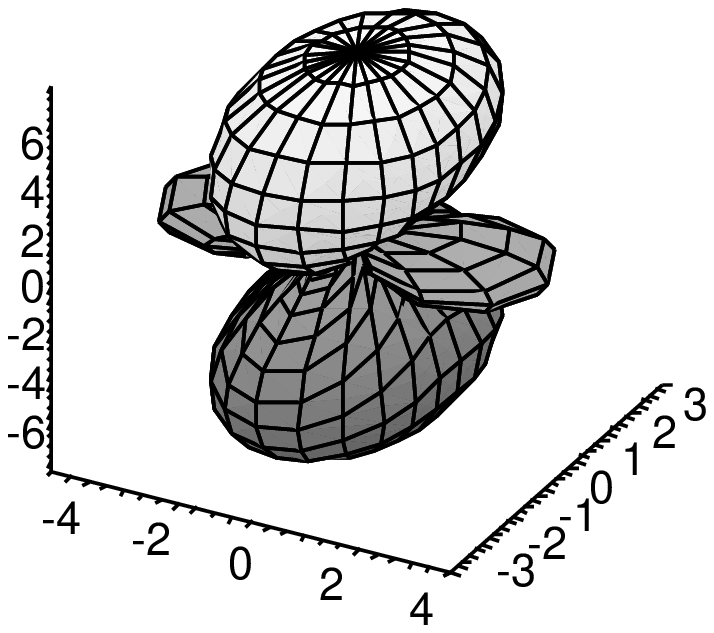}
    \end{center}
    \caption {A surface of quadrupole term of scalar potential,  $\vec {r}_0 \bot \vec 
{p}_0 $, $r_0 = 0.7$, $p_0 = 0.4$. }
\label{fig:6}
\end{figure}

Angular dependence of quadrupole term of scalar potential is:

\begin{equation}
\label{eq40}
\begin{array}{l}
D_{\alpha \beta } n^\alpha n^\beta =D_{zz} \cos ^2(\vartheta ) + \left( 
{D_{xx} \cos ^2(\varphi ) + D_{yy} \sin ^2(\varphi )} \right)\sin 
^2(\vartheta )+ \\
+ 2D_{xz} \cos (\vartheta )\sin (\vartheta )\cos (\varphi).
\end{array}
\end{equation}
This dependence can be the same as for the pair of classical electrons (Fig. \ref {fig:5}); the dependence could be changed to the inverse one (Fig. \ref {fig:4}); this dependence could loose it's axial symmetry (Fig. \ref {fig:6}).

\begin{figure}
\begin{center}
\includegraphics*[width=5cm]{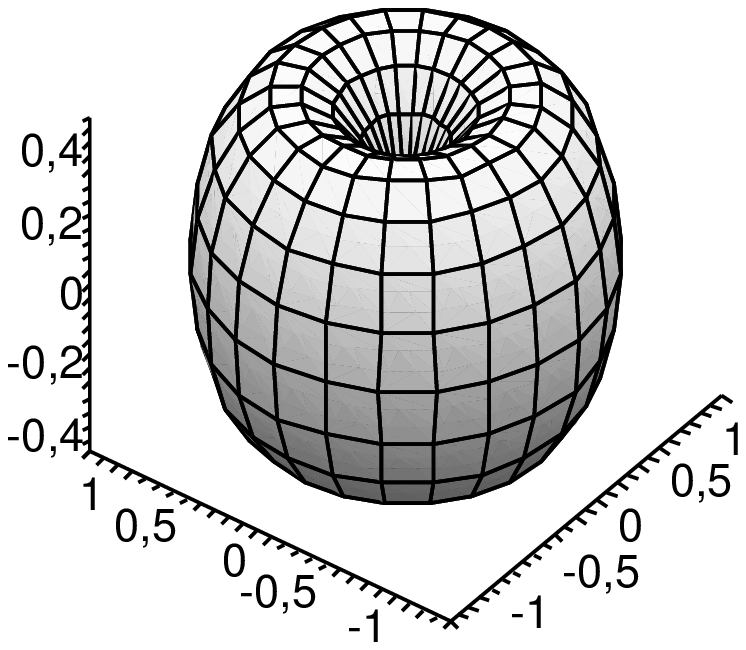}
\end{center}
\caption{A surface of quadrupole term of scalar potential,  $\vec {r}_0 \vert 
\vert \vec {p}_0 $, $r_0 = 0.42$, $p_0 = 0.3$.}
\label{fig:3}
\end{figure}

The calculation of quadrupole and magnetic moments of the coherent electron pair makes evidence of the following fact. In the case of small area of overlapping of wave functions the properties of moments are the same as for the the pair of classical charges. Otherwise the quadrupole moment is different. At the central impact of two coherent electrons the "light axis" is changed to "light plane", see Fig. \ref {fig:3} and Fig. \ref {fig:4}. At the non-central impact the axial symmetry of quadrupole term of scalar potential is lost, see Fig. \ref {fig:6}.

\section {Conclusions}

The results of the researches make evidence of the fact that relation between quadrupole electric and magnetic moments and relative coordinate and momentum is specific for the pair of quantum electrons, similar to classical case, while the coordinate and momentum uncertainties remain small comparatively with their values. 
The effect of magnetic moment is essential at small distances from the system of charges onlys. Deviations of quadrupole moment of coherent electron pair from a quadrupole moment for the pair of classical point charges are caused by coordinate uncertainty. One more cause of deviations is the interference of wave functions which depends on the relative momentum and its uncertainty. 
One of the most interesting implications of the existence of dependence of quadrupole moment on relative momentum is the loss of axial symmetry of quadrupole moment, as it is shown in Fig. \ref {fig:5}.
The influence of coordinate and momentum uncertainties is particularly evident in the nearest neighborhood of charges. Even in the case of separate electron, electrical and magnetic potentials do not tend to infinity. Instead, they are limited by the size of order $\frac{e_0 }{\sigma }$ (Fig. \ref {fig:2}). 

The dependence of quadrupole moment on relative momentum and relative coordinate makes it possible to calculate even small (smaller that or comparable to their uncertainties) values of relative coordinate and momentum. Relative coordinate and momentum have the meaning of observables.

\break
\appendix
\section{Definition of $Na(x)$}
\label{App}

Here we calculate $I = \int {\exp \left( { - \left( {\vec {r} - \vec {a}} 
\right)^2} \right)} \frac{d^3r}{\left| \vec {r} \right|}$, vector $\vec {a}$ 
is complex one, $\vec {a} = {\vec {a}}' + i{\vec {a}}''$. Here the direction 
of axis $z$ can not be chosen so as $z\left\| \vec {a} \right.$ that is why 
we considered the plane of vectors ${\vec {a}}'$ and ${\vec {a}}''$as 
coordinate plane.

\[
\int {\exp \left( { - \left( {\vec {r} - \vec {a}} \right)^2} \right)} 
\frac{d^3r}{\left| \vec {r} \right|} = \int {\exp \left( { - r^2 - a^2} 
\right)} \exp \left( {2\vec {r}\vec {a}} \right)\frac{d^3r}{\left| \vec {r} 
\right|},
\]

where

$\bar {r} = (r\sin \vartheta \cos \varphi ,r\sin \vartheta \sin \varphi 
,r\cos \vartheta ),\ 
\bar {a} = (a_x ,a_y ,0).$ A scalar product of these vectors is:

\[
\bar {r}\vec {a} = r\sin \vartheta \cos \varphi \cdot a_x + r\sin \vartheta 
\sin \varphi \cdot a_y = r\sin \vartheta \left( {a_ - e^{i\varphi } + a_ + 
e^{ - i\varphi }} \right).
\]

Then $I$ takes the following form:

\[I = \exp{ ( - a^2)}\int\limits_0^\infty {\int\limits_0^\pi{ 
\int\limits_0^{2\pi } {\exp{ ( - r^2)}   r\sin \vartheta \exp{ \left[ 
{2r\sin \vartheta \left( {a_ - e^{i\varphi } + a_ + e^{ - i\varphi }} 
\right)} \right]} d\varphi}d\vartheta}dr}.
\]

Now we expand exponent:

\[
I = \exp ( - a^2)\int\limits_0^\infty {\int\limits_0^\pi 
{\int\limits_0^{2\pi } {\exp ( - r^2)} } } r\sin \vartheta \sum\limits_{n = 
0}^\infty {\frac{(2r\sin \vartheta )^n}{n!}} \left( {a_ - e^{i\varphi } + a_ 
+ e^{ - i\varphi }} \right)^ndrd\vartheta d\varphi.
\]

Integrating over $\varphi $ we obtain:

\[
I_\varphi = \int\limits_0^{2\pi } {\left( {a_ - e^{i\varphi } + a_ + e^{ - 
i\varphi }} \right)^{2k}d\varphi } = 2\pi \left( {a_ - a_ + } 
\right)^{2k}\frac{(2k)!}{\left( {k!} \right)^2}.
\]

Thus, index $n$ is to take on only even values, $n = 2k$. 

\[
I = 2\pi \exp ( - a^2)\int\limits_0^\infty {\int\limits_0^\pi {\exp ( - 
r^2)r\sin \vartheta \sum\limits_{k = 0}^\infty {\frac{(2r\sin \vartheta 
)^{2k}}{\left( {2k} \right)!}\left( {a_ - a_ + } 
\right)^{2k}\frac{(2k)!}{\left( {k!} \right)^2}} ^{2k}drd\vartheta } } .
\]

Let us consider an integral:

\[
I_\vartheta = \int\limits_0^\pi {\sin \vartheta \sum\limits_{k = 0}^\infty 
{\frac{(2r\sin \vartheta )^{2k}}{\left( {k!} \right)^2}d\vartheta } } = 
2\sum\limits_{m = 0}^k {\frac{k!}{m!\left( {k - m} \right)!}} \frac{\left( { 
- 1} \right)^m}{\left( {2m + 1} \right)},
\]

summation over $r$ gives:

\[
I_r = \int\limits_0^\infty {\exp ( - r^2)r^{2k + 1}dr = \frac{1}{2}} k!
\]

Returning to the integral $I$we write:

\[
I = e^{ - a^2}I_\varphi I_\vartheta I_r = \pi \exp ( - a^2)\sum\limits_{k = 
0}^\infty {\sum\limits_{m = 0}^k {\left( {\left( \vec {a} \right)^2} 
\right)^k\frac{k!}{m!\left( {k - m} \right)!}} \frac{\left( { - 1} 
\right)^m}{\left( {2m + 1} \right)}} ,
\]

After sum over $m$ we obtain: 

\[
I = \pi ^{3 / 2}\exp ( - a^2)\sum\limits_{k = 0}^\infty 
{\frac{a^{2k}}{2\Gamma \left( {k + 3 / 2} \right)}} .
\]

Denotation we use:

\[
Na\left( {\vec {a}^2} \right) = \pi ^{3 / 2}\exp ( - a^2)\sum\limits_{k = 
0}^\infty {\frac{a^{2k}}{2\Gamma \left( {k + 3 / 2} \right)}} .
\]

\end{document}